\documentstyle[12pt]{article}
\makeatletter
\@addtoreset{equation}{section}
\makeatother

\newcommand{\be}{\begin{equation}}
\newcommand{\ee}{\end{equation}}
\newcommand{\bee}{\begin{eqnarray}}
\newcommand{\eee}{\end{eqnarray}}

\newcommand{\ga}{\alpha}
\newcommand{\gb}{\beta}
\newcommand{\gga}{\gamma}
\newcommand{\gd}{\delta}
\newcommand{\gl}{\lambda}

\newcommand{\gep}{\epsilon}
\newcommand{\gvep}{\varepsilon}
\newcommand{\gs}{\sigma}
\newcommand{\go}{\omega}

\newcommand{\half}{\frac{1}{2}}
\newcommand{\z}{\bar{z}}
\newcommand{\q}{\bar{q}}
\newcommand{\n}{\bar{n}}
\newcommand{\W}{\bar{W}}
\newcommand{\Y}{\bar{Y}}
\newcommand{\bxi}{\bar{\xi}}
\newcommand{\dt}{\partial_{\tau}}

\begin{document}

\thispagestyle{empty}

\begin{flushright}
\vspace{1mm}
FIAN/TD/16--99\\
{June 1999}\\
hep-th/9906149\\
\end{flushright}

\vspace{1cm}

\begin{center}
{\large\bf
CURRENTS OF ARBITRARY SPIN IN $AdS_3$}\\
\vglue 1  true cm
\vspace{1cm}
{\bf S.~F.~Prokushkin
\footnote{e-mail: prok@td.lpi.ac.ru}
  and M.~A.~Vasiliev }
\footnote{e-mail: vasiliev@td.lpi.ac.ru}  \\
\vspace{1cm}

I.E.Tamm Department of Theoretical Physics, Lebedev Physical
Institute,\\
Leninsky prospect 53, 117924, Moscow, Russia
\end{center}
\vspace{1cm}

\begin{abstract}
We study conserved currents of any
integer or half integer spin built from massless
scalar and spinor fields in $AdS_3$.
2-forms dual to the conserved currents in $AdS_3$ are
shown to be exact in the class of infinite
expansions in higher derivatives of the matter fields
with the coefficients containing inverse powers of the
cosmological constant. This property has no analog in the
flat space and may be related to the holography
of the AdS spaces.

\end{abstract}

\section{Introduction}\label{intro}

The role of anti-de Sitter (AdS) geometry in the high energy
physics increased greatly due to the Maldacena conjecture \cite{mald}
on the duality between the theory of gravity in the AdS space and
conformal theory on the boundary of the AdS space \cite{adscft,holo}.
The holography hypothesis suggests that the two types of
theories are equivalent. The same time, AdS geometry plays very
important role in the theory of higher spin (HS) gauge fields
(for a brief review see \cite{rev}) because interactions of HS
gauge fields contain negative powers of the cosmological
constant \cite{FV}. The theory of HS gauge fields may be
considered \cite{rev} as a candidate for a most symmetric phase of
string theory.

The group manifold case of $AdS_3$ is special and interesting
in many respects. HS gauge fields are not propagating in analogy
with the usual Chern-Simons gravitational and Yang-Mills fields,
although the HS gauge symmetries remain nontrivial.
The HS currents can be constructed from the
matter fields of spin 0 and spin 1/2. Their couplings to HS
gauge potentials describe interactions of the matter via
HS gauge fields.

Schematically, the equations of motion in the gauge field sector
have a form $R = J(C;W)$, where $R=dW-W\wedge W$ denotes all spin
$s\geq 1 $ curvatures built from the HS potential $W$, while $C$
denotes the matter fields (precise definitions are given in the
sect.~\ref{matter}). To analyze the problem perturbatively, one
fixes a vacuum solution $W_0$ that solves $R_0=0$, assuming that
$W=W_0 + W_1$, while $C$ starts from the first-order part.
When gravity is included, as is
the case in the HS gauge theories, $W_0$ is different from zero and
describes background geometry. In the lowest nontrivial order
one gets
\be
\label{R1}
    R_1 \equiv D_0 W_1 = J_2 (C^2) \,,
\ee
where $D_0$ is built from $W_0$ and the 2-form $J_2 (C^2)$
dual to the 3d conserved current vector field obeys
the conservation law
\be
\label{DJ2}
   D_0 J_2(C^2) = 0
\ee
on the free equations of motion of the matter fields.

A nonlinear system of equations of motion describing HS
gauge interactions for the spin 0 and spin 1/2 matter fields
in $AdS_3$ in all orders in interactions has been formulated both
for massless \cite{Eq} and massive \cite{PV} matter fields.
An interesting property of the proposed equations discovered
in \cite{PV} is that there exists a mapping of
the full nonlinear system to the free one. This mapping is
a nonlinear field redefinition having a form of infinite power
series in higher derivatives of the matter fields and is therefore
generically nonlocal. The coefficients of such expansions contain
inverse powers of the cosmological constant and therefore do not
admit a flat limit. We call such expansions in higher derivatives
pseudolocal. Comparison of the results of \cite{PV} with (\ref{R1})
implies that such a field redefinition
exists in a nontrivial model if
\be
\label{JDU}
   J_2 (C^2) =D_0 U (C^2) \,,
\ee
where $U$ is some pseudolocal functional of the matter fields.
The cohomological interpretation with $D_0$ as de Rahm differential
is straightforward because $D_0^2 = R_0 =0$.
Indeed, from (\ref{DJ2}) it follows that the current $J_2 (C^2)$
should be closed on the free equations for matter fields, while
(\ref{JDU}) implies that it is exact in the class of
pseudolocal functionals.
This fact has been already demonstrated for the spin 2 current in
\cite{PV}, where we have found a pseudolocal $U$ for the
stress tensor constructed from a massless  scalar field.
In this paper, we generalize this result to the currents of
an arbitrary integer or half integer spin.

Exact currents with local $U$ containing at most a finite
number of derivatives of the matter fields reproduce ``improvements'',
i.e., modifications of the currents which are trivially conserved.
The new result about AdS space established in this
paper is that the true currents can also be treated as ``improvements''
in the class of pseudolocal expansions. This sounds very suggestive
in the context of the holography hypothesis.

The paper is designed as follows. In sect.~\ref{matter} we collect
some facts about the equations of motion of the Chern-Simons HS
gauge fields and the ``unfolded'' formulation of the equations
of motion for the massless spin 0 and 1/2 matter fields in $AdS_3$.
In sect.~\ref{gener} we propose a formalism of generating functions
to describe differential forms bilinear in derivatives of the matter
fields. Then, in sect.~\ref{complex} we formulate using this method
the AdS on-mass-shell complex and study currents of
an arbitrary spin.

\section{Higher Spin and Matter Fields in $AdS_3$}\label{matter}

The 3d HS gauge fields are described \cite{Blen, Eq}
by a spacetime 1-form $W=dx^\mu W_\mu(y,\psi | x)$
depending on the spacetime coordinates $x^\mu$ ($\mu=0,1,2$),
auxiliary commuting spinor variables $y_{\ga}$
(indices $\ga\,,\,\gb\,,\gga\, =1,2$ are lowered and raised by
the symplectic form
$\gep_{\ga\gb}=-\gep_{\gb\ga}$, $\gep_{ 12} =\gep^{12}=1$,
$A^\ga =\gep^{\ga\gb}A_\gb$, $A_\ga = A^\gb \gep_{\gb\ga}$),
and the central involutive element $\psi$, 
\be
\label{W}
  W_\mu (y,\psi | x) = \sum_{n=0}^\infty {1\over 2in!} \;
  \left[\go_{\mu,\;\ga(n)} (x) + \gl\psi \;
  h_{\mu,\;\ga(n)} (x) \right] \;
  y^{\ga_1}\ldots y^{\ga_n}  \,.
\ee
A constant parameter $\gl$ is to be identified with
the inverse radius of $AdS_3$.

The HS gauge algebra is a Lie superalgebra
built via (anti)commutators from the associative algebra
spanned by the elements $a(y,\psi)$ with the product law ``$*$''
defined via the generating relations
$y_\ga * y_\gb - y_\gb * y_\ga = 2i\epsilon_{\ga\beta}$,
$y_\ga * \psi = \psi * y_\ga$, $\psi * \psi = 1$.
(The boson-fermion parity $\pi$ is defined in a usual way as
$a(-y,\psi) = (-)^{\pi(a)}\, a(y,\psi)$.)
The field strength is \cite{HS, Blen}
$R(y,\psi | x) = dW(y,\psi | x) -
W(y,\psi | x) * \wedge W(y,\psi | x)$.
The role of the element $\psi$ is to make the 3d HS
superalgebra semisimple ($hs(2)\oplus hs(2)$, where, in notation of
\cite{HS}, $hs(2)$ is a superalgebra spanned by the $\psi$-independent
elements $a(y)$), with simple components singled out by the projectors
$P_\pm = \half (1\pm\psi )$. This is similar to
the $AdS_3$ isometry algebra $o(2,2)\sim sp(2)\oplus sp(2)$.
The latter is identified with
a subalgebra of $hs(2)\oplus hs(2)$ spanned by
$L_{\ga\gb}= {1\over 2i} \; y_\ga y_\gb$ and
$P_{\ga\gb} = {1\over 2i} \; y_\ga y_\gb \psi$.
We therefore identify the $o(2,2)$ components of $W(y,\psi | x)$
(\ref{W}) with the gravitational Lorentz connection 1-form
$\go^{\ga\gb}(x)=dx^\mu\go_{\mu,}{}^{\ga\gb}(x)$ and the dreibein
1-form $h^{\ga\gb}(x)=dx^\mu h_{\mu,}{}^{\ga\gb}(x)$. Since
$AdS_3$ algebra $o(2,2)$ is a proper subalgebra of the d3 HS
algebra it is a consistent ansatz to require the vacuum value of
$W(y,\psi | x)$ to be non-zero only in the spin 2 sector. Then
the equation $R_0 = 0$ is equivalent to the $o(2,2)$ zero-curvature
conditions
\be
\label{d omh}
    d\go_{\ga\gb}=\go_{\ga\gga}\wedge\go_\gb{}^\gga
    + \gl^2h_{\ga\gga}\wedge h_\gb{}^\gga   \,,\qquad
    dh_{\ga\gb}=\go_{\ga\gga}\wedge h_\gb{}^\gga
    + \go_{\gb\gga}\wedge h_\ga{}^\gga \,.
\ee
For the metric interpretation, the dreibein $h_{\nu,}{}^{\ga\gb}$
should be non-degenerate, thus admitting the inverse dreibein
$h^{\nu}{}_{,\,\ga\gb}$ defined via
$h_{\nu,}{}^{\ga\gb} h^{\nu}{}_{,\,\gga\gd}=
\frac12\,(\gd^\ga_\gga\gd^\gb_\gd+\gd^\ga_\gd\gd^\gb_\gga)$.
Then, the second
equation in (\ref{d omh}) reduces to the zero-torsion condition which
expresses Lorentz connection $\go_{\nu,}{}^{\ga\gb}$ via dreibein
$h_{\nu,}{}^{\ga\gb}$, while the first one implies that
${\cal R}_{\ga\gb}= -\gl^2h_{\ga\gga}\wedge h_\gb{}^\gga$, where
${\cal R}_{\ga\gb}$ is the Riemann tensor 2-form. Therefore,
the equations (\ref{d omh}) describe $AdS_3$ with radius $\gl^{-1}$.

The massless Klein-Gordon and Dirac equations in $AdS_3$ read
\be
\label{K-G D}
  \Box C = \frac32\gl^2 \; C \qquad   
  \mbox{and} \qquad
  h^{\mu}{}_{,\,\ga}{}^\gb \nabla_{\mu}C_\gb = 0
\ee
for the spin 0 boson field $C(x)$ and spin $1\over 2$ fermion field
$C_{\ga}(x)$. Here $\Box =\nabla^{\mu}\nabla_{\mu}$, where
$\nabla_{\mu}$ is the full covariant derivative with
the symmetric Christoffel connection defined via the metric postulate
$\nabla_{\mu} h_{\nu,}{}^{\ga\gb}=0$.

The ``unfolded'' formulation \cite{Unf} of the equations
(\ref{K-G D}) in the form of some covariant constancy conditions
is most convenient for the analysis of cohomology of currents.
To this end one introduces an infinite set of symmetric multispinors
$C_{\ga_1\ldots\ga_{n}}$ for all $n\ge 0$.
(We will assume total symmetrization of
indices denoted by the same letter and will use the notation
$C_{\ga(n)}=C_{\ga_1 \dots \ga_n}$ when only a number of indices is
important.) As shown in \cite{Unf}, the infinite chain of equations
\be
\label{chain}
   D^L C_{\ga(n)} = {i\over 2} \left[ h^{\gb\gga}C_{\gb\gga\ga(n)}
    - \gl^2 n(n-1) \; h_{\ga\ga} C_{\ga(n-2)} \right] \,,
\ee
where $D^L$ is the background Lorentz covariant differential,
$D^L C_{\ga(n)}=dC_{\ga(n)}+n\; \go_{\ga}{}^\gga C_{\gga\ga(n-1)}$,
is equivalent to the equations (\ref{K-G D}) for the lowest rank
components $C$ and $C_{\ga}$ along with some constraints expressing
highest multispinors via highest spacetime derivatives of $C$ and
$C_{\ga}$. For example, for bosons
\be
\label{der}
   C_{\ga (2n)}(x) =
   (-2i)^n\; h^{\nu_1}{}_{,\,\ga\ga} h^{\nu_2}{}_{,\,\ga\ga}
   \ldots h^{\nu_n}{}_{,\,\ga\ga}\;
   \nabla_{\nu_1} \nabla_{\nu_2}\ldots \nabla_{\nu_n} \; C(x) \,,
\ee
where $\nabla_{\mu}$ is a full background derivative (for multispinors
$\nabla_{\mu} C_{\ga(n)} = D^L_{\mu} C_{\ga(n)}$).

Following \cite{Unf}, let us introduce the generating function
\be
\label{Cy}
  C(y,\psi | x) = \sum_{n=0}^\infty
  {1\over n!}\; (\gl^{-1}\psi)^{[\frac n2]}  \;
  C_{\ga_1 \ldots \ga_n}(x) \;
  y^{\ga_1}\ldots y^{\ga_n} = \gl^{\half\pi (C)} \;
  \tilde{C}(\gl^{-\half} y, \psi | x)\,,
\ee
where $[n + a] = n$, $\forall n \in Z$ and $0\leq a <1$, and
the boson-fermion parity $\pi (C)=0(1)$ for even (odd) functions
$C(y)$. The equations (\ref{chain}) can be rewritten
in the form \cite{Unf},
\be
\label{DC}
  D^L C(y,\psi) = {i\gl \over 2}\,\psi \; h^{\ga\gb}
  \left[\frac{\partial}{\partial y^\ga }
  \frac{\partial}{\partial y^\gb}
  - y_\ga y_\gb \right] \; C(y,\psi) \,,\qquad
  D^L = d - \go^{\ga\gb}y_\ga \frac{\partial}{\partial y^\gb} \,.
\ee
The fields $C_{\ga_1 \ldots \ga_n}$ are identified with all
on-mass-shell nontrivial
derivatives of the matter fields according to (\ref{der}).
The condition that the system is on--mass--shell is encoded in the
fact that the multispinors $C_{\ga_1 \ldots \ga_n}$ are totally
symmetric. This allows us to work with $C_{\ga_1 \ldots \ga_n}$
instead of explicit derivatives of the matter fields.
Consider a function $F[C_{\ga (n)}(x)]$ of all components of
$C_{\ga_1 \ldots \ga_n}(x)$ at some fixed point $x$. $F$ is
not supposed to contain any derivatives with respect to the spacetime
coordinates $x$ and therefore looks like a local function of matter
fields. One has to be careful however because, when the equations
(\ref{chain}) hold, (\ref{der}) is true. We will therefore call
a function $F[C_{\ga (n)}]$ pseudolocal if it is an infinite
expansion in the field variables $C_{\ga (n)} (x)$ and local if
$F$ is a polynomial. In terms of the generating functions
$C(y,\psi |x)$ this can be reformulated as follows.
Let $F(C|x)$ be some functional
of the generating function $C(y,\psi | x)$ at some fixed point of
spacetime $x$. According to (\ref{der}) its spacetime locality is
equivalent on--mass--shell to the locality in the $y$ space.
Indeed, from (\ref{DC}) it follows that the derivatives in the spinor
variables form in a certain sense a square root of the spacetime
derivatives.

The equation (\ref{R1}) for the d3 HS system reads
(in the rest of the paper we use the symbol $D$ instead of $D_0$)
\be
\label{Ch-S}
   D W_1(y,\psi | x) = J(C^2) (y,\psi | x)
\ee
with the background AdS covariant differential
\be
\label{D}
   D = D^L - \gl\psi\; h^{\ga\gb} \;
   y_{\ga}\frac{\partial}{\partial y^\gb} = d - (\go^{\ga\gb}
   + \gl\psi\; h^{\ga\gb}) \;
   y_{\ga}\frac{\partial}{\partial y^\gb} \,.
\ee
That $\go^{\ga\gb}(x)$ and $h^{\ga\gb}(x)$ obey the equations
(\ref{d omh}) guarantees $D^2 =0$.
Thus, our problem is to study the cohomology of $D$ (\ref{D}).
Clearly, $D$ commutes with the
Euler operator $N=y^\ga \frac{\partial}{\partial y^\ga}$.
Its eigenvalues are identified with spin $s$ via $N = 2(s-1)$.
The problem therefore is to be analyzed for different spins
independently.

Conserved currents of an arbitrary integer spin in d4 Minkowski
spacetime were considered in \cite{BBD}. For $d=2$, HS
conserved currents were constructed in \cite{BB}.

In the case of $AdS_3$, conserved currents of any integer spin $s\ge 1$
built from two massless scalar fields $C$, $C'$ have a form
\bee
\label{s-scal}
    \hspace*{-1.5cm}
    J^{(s)}_{\mu,\,\ga(2s-2)}(C,C') &=& \sum_{k=0}^{s-2}\;
    {2(-1)^k\over (2k+1)!(2s-2k-3)!}\;
    h_{\mu,}{}^{\gga\gga}\;
    C_{\gga\ga(2k+1)} C'_{\gga\ga(2s-2k-3)} \nonumber\\
    &&\hspace*{-3.5cm}{}+\sum_{k=0}^{s-1} {(-1)^k\over (2k)!(2s-2k-2)!}\;
    h_{\mu,}{}^{\gga\gga}\;
    \left[ C_{\gga\gga\ga(2k)} C'_{\ga(2s-2k-2)}
    - C_{\ga(2k)} C'_{\gga\gga\ga(2s-2k-2)} \right] \,.
\eee
Analogously, one can write down the currents of any integer spin
$s\ge 1$ built from two massless spinors, and the ``supercurrents''
of any half integer spin $s\ge 3/2$ built from one scalar and one
spinor \cite{fut}. The lowest spin conserved currents read
\bee
\label{s=1}
   && \hspace*{-2cm}
   J_{\mu}^{(1)}(C,C') = h_{\mu,}{}^{\gga\gga}\;
   (C_{\gga\gga} C' - C C'_{\gga\gga})  \,, \qquad
   J_{\mu}^{(1)}(C_\ga,C'_\ga) = h_{\mu,}{}^{\gga\gga}\;
   C_\gga C'_\gga \,, \\
\label{s=3/2}
   && \hspace*{-2cm}
   J_{\mu,\,\ga}^{({3/2})}(C,C'_\ga) = h_{\mu,}{}^{\gga\gga}\;
   (C_{\gga\gga} C'_\ga - C C'_{\gga\gga\ga}
   + 2 C_{\gga\ga} C'_\gga) \,, \\
\label{s=2}
   && \hspace*{-2cm}
   J_{\mu,\,\ga\ga}^{(2)}(C,C') =
   \half\, h_{\mu,}{}^{\gga\gga}\;
   (C_{\gga\gga} C'_{\ga\ga} - C C'_{\gga\gga\ga\ga}
   - C_{\gga\gga\ga\ga} C' + C_{\ga\ga} C'_{\gga\gga}
   + 4\, C_{\gga\ga} C'_{\gga\ga})  \,.
\eee
These currents are all local, containing a finite number of terms
(i.e., higher derivatives (\ref{der})). The same expressions remain
valid in the flat limit with $\nabla_\mu \to \partial_\mu$
in (\ref{der}).

\section{Generating Functions}\label{gener}

To analyze the cohomology problem for currents of an arbitrary spin
we first elaborate a technique operating with the generating
functions (\ref{Cy}) rather than with the individual multispinors.
A generic Lorentz covariant spacetime 1-form of spin $s=n/2+1$
bilinear in two different matter fields $C$ and $C^\prime$ and
their on--mass--shell nontrivial derivatives is
\bee
\label{genform}
    \Phi_{\ga (n)}\,(C,C'|x) &=&
    \sum_{k+l=n-2} \sum_{m=0}^\infty a^{klm}\;
    h_{\ga\ga} C_{\ga(k)}{}^{\gb(m)}(x)\; C'_{\ga(l)\gb(m)}(x)
    \nonumber \\
    &&\hspace*{-3.5cm}{}+ \sum_{k+l=n-1} \sum_{m=0}^\infty
    \left[ b_1^{klm}\; h_\ga{}^\gga
    C_{\gga\ga(k)}{}^{\gb(m)}(x)\; C'_{\ga(l)\gb(m)}(x)
    + b_2^{klm}\; h_\ga{}^\gga C_{\ga(k)}{}^{\gb(m)}(x)\;
    C'_{\gga\ga(l)\gb(m)}(x) \right]   \nonumber \\
    &&\hspace*{-3.5cm}{}+ \sum_{k+l=n} \sum_{m=0}^\infty
    \left[ e_1^{klm}\;h^{\gga\gga} C_{\gga\gga\ga(k)}{}^{\gb(m)}(x)\;
    C'_{\ga(l)\gb(m)}(x)+ e_2^{klm}\; h^{\gga\gga}
    C_{\gga\ga(k)}{}^{\gb(m)}(x)\;
    C'_{\gga\ga(l)\gb(m)}(x)  \right. \nonumber \\
    &&\hspace*{-3.5cm}\left.{}+ e_3^{klm}\; h^{\gga\gga}
    C_{\ga(k)}{}^{\gb(m)}(x)\;
    C'_{\gga\gga\ga(l)\gb(m)}(x)  \right] \,,
\eee
where $a^{klm}$, $b_{1,2}^{klm}$, and $e_{1,2,3}^{klm}$ are
arbitrary constants and $h_{\ga\ga}$ is the dreibein 1-form.
Introducing
$\Phi(y,\psi | x) = \Phi_{\ga_1 \ldots \ga_n}(\psi | x)\,
y^{\ga_1}\ldots y^{\ga_n}$,
one can equivalently rewrite this formula as
\bee
\label{int1}
  \Phi(y,\psi | x) &=& h_{\ga\ga} \;
    {1\over (2\pi)^2} \oint dr  \oint ds
    \oint \tau^{-2}\,d\tau \int d^2 u \, d^2 v \;
    \exp\left\{{i \over \tau} (u_\gga v^\gga)\right\} \nonumber \\
  &&\hspace*{-2cm} {}\times C(u-ry,\psi | x)\; C'(v+sy,\psi | x)
    \left[ f_1(r,s,\tau)\, y^{\ga} y^{\ga}
    + f_2(r,s,\tau)\, y^{\ga} u^{\ga}
    + f_3(r,s,\tau)\, y^{\ga} v^{\ga} \right. \nonumber \\
  &&\hspace*{-2cm}\left. {}+ f_4(r,s,\tau)\, u^{\ga} u^{\ga}
    + f_5(r,s,\tau)\, u^{\ga} v^{\ga}
    + f_6(r,s,\tau)\, v^{\ga} v^{\ga} \right] \,.
\eee
Here $r$, $s$, and $\tau$ are complex variables, $u_{\ga}$ and
$v_{\ga}$ $(\ga=1,2)$ are spinor variables. The quantities
$f_i(r,s,\tau)$, $i=1, \ldots, 6$ are polynomials in $r^{-1}$
and $s^{-1}$ and formal series in $\tau^{-1}$,
\be
\label{fi}
    f_i(r,s,\tau) = \sum_{0<k,\,l<p, \atop p<\infty} \sum_{m=1}^\infty
    f_i(k,l,m)\; r^{-k} s^{-l} \tau^{-m}   \,.
\ee
The contour integrations are normalized as
$\oint \tau^{-n}d\tau = \gd_n^1$. The Gaussian integrations
with respect to $u_\ga$ and $v_\ga$ should be completed prior
the contour integrations.

Inserting (\ref{Cy}) into (\ref{int1}) and completing elementary
integrations one arrives at (\ref{genform}) with the coefficients
$a^{klm}$, $b_{1,2}^{klm}$, and $e_{1,2,3}^{klm}$ expressed via
$f_i(k,l,m)$. For example,
\be
\label{a}
    a^{klm} = {(-)^{k+m}i^m\over k!\, l!\, m!}
    (\gl^{-1}\psi)^{\left[{k+m\over 2}\right]
    + \left[{l+m\over 2}\right]}\; f_1(k+1,\,l+1,\,m+1) \,.
\ee
Therefore (\ref{int1}) indeed describes a general Lorentz covariant
1-form bilinear in the matter fields. Note that the formula
(\ref{int1}) produces a spacetime local expression if all
the coefficients $f_i$ contain a finite number of terms in
(\ref{fi}) and pseudolocal if some of the expansions in negative
powers of $\tau$ are infinite.

In practice, the following representations of rank $n=0,1,2,3$
differential forms $\Phi_n(x)$ are most convenient,
\bee
\label{0-form}
    \hspace*{-0.8cm}
    \Phi_{0,3}(y,\psi|x) &=& H_{0,3}\; {1\over (2\pi)^2}
    \oint {dz\over z} \oint {d\z\over\z}
    \oint {d\tau\over \tau^2} \int d^2 q\,d^2 \q\;
    \exp\left\{-{1 \over 2\tau} (q_\gga\q^\gga)\right\}   \nonumber \\
    &&\hspace*{-3.2cm}{} \times
    C\left[ \left.\half(q+\q) - {1\over 2i}(z-\z)y\,,\;
    \psi\,\right| \, x \right]
    C'\left[ \left.{1\over 2i}(q-\q) + \half(z+\z)y\,,\;
    \psi\,\right| \,x\right] E_{0,3}(z, \z, \tau) \,, \\
\label{int}
   \hspace*{-0.8cm}
   \Phi_{1,2}(y,\psi|x) &=& H_{1,2\; \ga\ga} \;
    {1\over (2\pi)^2}\oint dz  \oint d\z
    \oint \tau^{-2}\,d\tau \int d^2 q \, d^2 \q\;
    \exp\left\{-{1 \over 2\tau} (q_\gga \q^\gga)\right\} \nonumber \\
  &&\hspace*{-2.5cm}
    \times C\left[ \left.\half(q+\q) - {1\over 2i}(z-\z)y\,,\;
    \psi\,\right| \,x \right]
    C'\left[ \left.{1\over 2i}(q-\q) + \half(z+\z)y\,,\;
    \psi\,\right| \,x \right]   \nonumber \\
  &&\hspace*{-2.5cm} {}\times\left\{R_{1,2}(z,\z,\tau)\, y^{\ga} y^{\ga}
    +{1\over 2\tau \z} W_{1,2}(z,\z,\tau)\, y^{\ga} \q^{\ga}
    +{1\over 2\tau z} \W_{1,2}(z,\z,\tau)\, y^{\ga} q^{\ga}
    \right. \nonumber \\
  &&\hspace*{-2.5cm}\left.
    {}+{1\over 2\tau^2 \z^2} Y_{1,2}(z,\z,\tau)\,\q^{\ga} \q^{\ga}
    +{1\over 2\tau^2 z^2} \Y_{1,2}(z,\z,\tau)\, q^{\ga} q^{\ga}
    +{1\over 2\tau^2 z\z} V_{1,2}(z,\z,\tau)\, q^{\ga} \q^{\ga}
    \right\} \,.
\eee
Here $H_0=\psi$,
$H_3=-{\gl^2\over 12}\,
h_{\ga\gb} \wedge h^{\gb}{}_{\gga} \wedge h^{\gga\ga}$,
$H_{1\; \ga\ga}=h_{\ga\ga}$, and
$H_{2\; \ga\ga}=-{\gl\over 2}\psi\,
h_{\ga\gb} \wedge h^{\gb}{}_{\ga}$, where the factors of $\psi$,
$-{\gl\over 2}\psi$, and $-{\gl^2\over 12}$ are introduced for future
convenience. It is not hard to see that the expressions (\ref{0-form})
and (\ref{int}) reproduce arbitrary Lorentz covariant forms bilinear
in the matter fields and their on-mass-shell nontrivial derivatives.

Let $n$, $\n$, and $n_\tau$ be the following operators,
\be
\label{nn}
   n = z\, {\partial\over \partial z} \,,\qquad
   \n = \z\, {\partial\over \partial \z} \,,\qquad
   n_\tau = \tau\, {\partial\over \partial\tau}
\ee
(using the same notations for their eigenvalues).
The quantities $R_{1,2}(z,\z,\tau)$, $W_{1,2}(z,\z,\tau)$,
$\W_{1,2}(z,\z,\tau)$, $Y_{1,2}(z,\z,\tau)$,
$\Y_{1,2}(z,\z,\tau)$, $V_{1,2}(z,\z,\tau)$, and
$E_{0,3}(z,\z,\tau)$ give a non-zero contribution to (\ref{0-form})
and (\ref{int}) when $n$, $\n$, and $n_\tau$ satisfy the following
restrictions:
\be
\label{nonzero}
\begin{array}{|l|ccr|ccr|ccr|}
\hline
\vphantom{\displaystyle{\hat{\W}}}
R_1,\, R_2\hspace{1cm}
&\quad n&\le &-1 \quad &\quad\n &\le &-1 \quad
&\quad n_\tau &\le &-1 \quad\\
\hline
\vphantom{\displaystyle{\hat{\W}}}
W_1,\, W_2\hspace{1cm}
&\quad n&\le &-1 \quad &\quad \n &\le & 0 \quad
&\quad n_\tau &\le &-1 \quad\\
\hline
\vphantom{\displaystyle{\hat{\W}}}
\W_1,\, \W_2\hspace{1cm}
&\quad n&\le & 0 \quad &\quad\n &\le &-1 \quad
&\quad n_\tau &\le &-1 \quad\\
\hline
\vphantom{\displaystyle{\hat{\W}}}
Y_1,\, Y_2\hspace{1cm}
&\quad n&\le &-1 \quad &\quad\n &\le & 1 \quad
&\quad n_\tau &\le &-1 \quad\\
\hline
\vphantom{\displaystyle{\hat{\W}}}
\Y_1,\, \Y_2\hspace{1cm}
&\quad n&\le & 1 \quad &\quad\n &\le &-1 \quad
&\quad n_\tau &\le &-1 \quad\\
\hline
\vphantom{\displaystyle{\hat{\W}}}
V_1,\, V_2,\, E_0,\, E_3
&\quad n&\le & 0 \quad &\quad \n &\le & 0 \quad
&\quad n_\tau &\le &-1 \quad\\
\hline
\end{array}
\ee
Beyond these regions, the coefficients do not contribute
and therefore their values can be fixed arbitrarily.
As a result, the quantities $R_{1,2}$, $W_{1,2}$, $\W_{1,2}$, ...
are defined modulo arbitrary polynomials
$P(\tau) = \sum_{k=0}^{k_0} P_k \tau^k$.

For the two-component spinors, antisymmetrization over any three
two-component spinor indices gives zero. This is expressed by
the identity
$a_\ga(b_\gb c^\gb)+b_\ga(c_\gb a^\gb)+c_\ga(a_\gb b^\gb)=0$
valid for any three commuting two-component spinors $a_\ga$, $b_\ga$,
and $c_\ga$. As a result, the forms discussed so far are not all
independent. The ambiguity in adding any terms which vanish as
a consequence of this identity can be expressed in a form of some
equivalence (gauge) transformations of the coefficients in
(\ref{int1}) and (\ref{int}).
We call these equivalence transformations Fierz transformations.
Using the partial integrations w.r.t. $\tau$, $z$, and $\z$, one
can check that the transformations
\bee
\label{delta}
   &&\hspace*{-1.5cm}\delta R_{1,2} = -\dt \chi_{1,2} \,,\qquad
   \delta W_{1,2} = -\dt  \xi_{1,2} + 2i\n \chi_{1,2} \,,\qquad
   \delta \W_{1,2} = -\dt \bxi_{1,2} - 2in \chi_{1,2} \,,
   \nonumber\\
   &&\hspace*{-1.5cm}
   \delta V_{1,2} = -i(n\xi_{1,2} - \n\bxi_{1,2})\,, \qquad
   \delta Y_{1,2} = i(\n-1)\xi_{1,2}  \,, \qquad
   \delta \Y_{1,2} = -i(n-1)\bxi_{1,2}
\eee
with arbitrary parameters $\chi_{1,2}(z,\z,\tau)$,
$\xi_{1,2}(z,\z,\tau)$, and $\bxi_{1,2}(z,\z,\tau)$
describe all possible Fierz transformations
of the 1- and 2-forms (\ref{int}).

\section{On-Mass-Shell Current Complex}\label{complex}

In this section we study the on--mass-shell action of the operator
$D$ (\ref{D}) on the differential forms defined in sect.~\ref{gener}.
The advantage of the formulation of the dynamical equations
in the unfolded form (\ref{DC}) is that it expresses the spacetime
derivative of $C$ via some operators acting in the auxiliary spinor
space. As a result, on--mass--shell action of $D$ reduces to some
mapping ${\cal D}$ acting on the coefficients in the formulae
(\ref{0-form})-(\ref{int}).

Let us consider the example of a 0-form. Using the Leibnitz rule for
$D^L$ and taking into account the equations of motion (\ref{DC}) and
the zero torsion condition $D^L h_{\ga\ga} = 0$ (\ref{d omh}), one
arrives at the 1-form $\hat{\Phi}_1 = D \Phi_0$ with the coefficients
$R_1^{\cal D} (E_0)$, $W_1^{\cal D} (E_0)$, $\W_1^{\cal D}(E_0)$,
$Y_1^{\cal D}(E_0)$, $\Y_1^{\cal D}(E_0)$, and $V_1^{\cal D}(E_0)$
of the form
\bee
\label{^phi1}
  && R_1^{\cal D} (E_0) = - {i\gl\over 2} E_0\,,\qquad
  W_1^{\cal D} (E_0) = i\gl(1-i\tau) E_0\,,\qquad
  \W_1^{\cal D} (E_0) = i\gl(1+i\tau) E_0 \,, \nonumber \\
  && V_1^{\cal D} (E_0) = -i\gl(1+\tau^2) E_0 \,, \qquad
  Y_1^{\cal D} (E_0) = \Y_1^{\cal D} (E_0) = 0 \,.
\eee
Analogously one derives the mapping
$\left.{\cal D}:\; \Phi_i(y)\to \hat{\Phi}_{i+1}(y)
= D \Phi_i(y)\right|_{on-shell}$, $i=1,2$ on the coefficients of
the differential forms (\ref{0-form}), (\ref{int}),
${\cal D}\; \{R_1, W_1, ... \} = \{R_2^{\cal D}, W_2^{\cal D}, ... \}$,
${\cal D}\; \{R_2, W_2, ... \} = E_3^{\cal D}$ with
\bee
\label{^R}
   R_2^{\cal D} &=&
   -(1-i\tau)nR_1 - (1+i\tau)\n R_1 + 2R_1
   +{i\over 4}(1+i\tau)\dt W_1
   -{i\over 4}(1-i\tau)\dt \W_1   \nonumber \\
   & & {}+{1\over 4}(W_1+\W_1)-\half(nW_1+\n\W_1) \,,  \\
\label{^W}
   W_2^{\cal D} &=& -{i\over 2}\dt [(1+\tau^2)W_1]
   +{3\over 2}(1+i\tau)W_1 + 2(1+\tau^2)\n R_1
   +\half (1-i\tau)\n\W_1 + (1-2n)Y_1   \nonumber \\
   & & {}-\half (1+i\tau)(\n-1)W_1
   +i(1+i\tau)\dt Y_1+\left({3\over 2}-\n \right) V_1
   -{i\over 2}(1-i\tau)\dt V_1  \,, \\
\label{^V}
   V_2^{\cal D} &=& \half (1+\tau^2)(nW_1+\n\W_1)
   +\half (1+i\tau)(\n-1) V_1
   +\half (1-i\tau)(n-1)V_1 + V_1  \nonumber \\
   & & {}+(1+i\tau) nY_1 +(1-i\tau)\n\Y_1  \,, \\
\label{^Y}
   Y_2^{\cal D} &=& \half (1+\tau^2)(\n-1) W_1
   -i(1+\tau^2)\dt Y_1
   +(1+i\tau)Y_1 + (1-i\tau) nY_1  \nonumber \\
   & & {}+\half (1-i\tau)(\n-1) V_1  \,,  \\
\label{^G}
    E_3^{\cal D} &=&   4i n\n (1+\tau^2) R_2\,
    + in\n (W_2+\W_2) - 3\tau (nW_2-\n\W_2) - \tau n\n (W_2-\W_2)
    \nonumber\\
   && {} + \dt [(1+\tau^2)(nW_2-\n\W_2)]
    + 2\dt (nY_2-\n\Y_2) - (n-\n)\;\dt  V_2 \nonumber\\
   && {} + 2i(n\n-n-\n+1) V_2
    + i(n+\n-2)\;\tau\dt  V_2 + i(1+\tau^2)\;\dt\dt V_2 \nonumber\\
   && {} - 3i (nW_2+\n\W_2) + (i+\tau)\; n(n+1) W_2
    + (i-\tau)\;\n(\n+1) \W_2   \nonumber\\
   && {} - 4i (nY_2+\n\Y_2) + 2i \;
   \tau\dt  (nY_2+\n\Y_2) + 2i [n(n+1) Y_2 + \n(\n+1) \Y_2]  \,.
\eee
The corresponding formulae for the parameters $\W_2^{\cal D}$
and $\Y_2^{\cal D}$ are given by (\ref{^W}) and (\ref{^Y}) respectively
with the replacements $i\to -i$, $n\leftrightarrow\n$,
$W_1\leftrightarrow\W_1$, and $Y_1\leftrightarrow\Y_1$ on r.h.s.-s.
As expected, ${\cal D}^2 = 0$ and therefore the mapping ${\cal D}$
defines a complex $(T, {\cal D})$ with
$T = \bigoplus\limits_{i=0,1,2,3} T_i$, where
$T_{0,3} = \{E_{0,3} \}$ and
$T_{1,2} = \{R_{1,2}, W_{1,2}, \W_{1,2}, V_{1,2},
Y_{1,2}, \Y_{1,2}\}$. The reformulation of the problem in terms of
$(T, {\cal D})$ effectively accounts the fact that the fields are
on--mass--shell. We identify the cohomology of currents with
the cohomology of ${\cal D}$.

The remarkable property of the mapping ${\cal D}$ is that
it contains $z$, $\z$, ${\partial\over\partial z}$, and
${\partial\over\partial \z}$ only via $n$ and $\n$ (\ref{nn}),
thus implying the separation of variables: the differential
${\cal D}$ leaves invariant eigensubspaces of $n$ and $\n$.
In fact, this is the main reason for using the particular
represenation (\ref{0-form})-(\ref{int}).

As expected, the system (\ref{^R})-(\ref{^Y}) is consistent
with the Fierz transformations (\ref{delta}).
Namely, any Fierz transformation of the quantities $R_1$, $W_1$, ...
leads to some Fierz transformation of the quantities $R_2^{\cal D}$,
$W_2^{\cal D}$, ... , and any Fierz transformation of $R_2$,
$W_2$, ... does not affect the parameter $E_3^{\cal D}$ (\ref{^G}).

Following \cite{BBD} we study the currents containing
the minimal possible number of spacetime derivatives for a given
spin $s$. {}From (\ref{der}) it is clear that this is the case
if the number of the contracted indices $\gb$ in $(\ref{genform})$
is zero. Since the number of contractions is $-(n_\tau+1)$
(see sect.~\ref{gener}) we consider 2-forms
$\Phi_2^{n,\n}$ with $n_\tau = -1$. Thus we set in (\ref{int})
\bee
\label{a:t}
   &&\hspace*{-1cm} R_2=\ga_R(n,\n)\; z^n \z^{\n} \tau^{-1}\,,
   \quad
   W_2=\ga_W(n,\n)\; z^n \z^{\n} \tau^{-1}\,, \quad
  \W_2=\ga_{\W}(n,\n)\; z^n \z^{\n} \tau^{-1}\,, \nonumber \\
   &&\hspace*{-1cm} Y_2=\ga_Y(n,\n)\; z^n \z^{\n} \tau^{-1} \,,
   \quad
  \Y_2=\ga_{\Y}(n,\n)\; z^n \z^{\n} \tau^{-1}\,,\quad
   V_2=\ga_V(n,\n)\; z^n \z^{\n} \tau^{-1} \,
\eee
with some constant parameters
$\ga_R(n,\n),\, \ga_W(n,\n),\, ... \sim \gl^{[s]}$, where
$s=1-\half(n+\n)$. The conservation condition means that
$\Phi_2^{n,\n}$ should be ${\cal D}$-closed. The requirement
$E_3^{\cal D} = 0$ modulo terms that do not contribute to
(\ref{0-form}) imposes the following conditions
\bee
\label{closed}
   &&\hspace*{-2cm} 4n\n \ga_R + (n+\n-2)(n\ga_W + \n\ga_{\W})
   + 2n(n-2) \ga_Y  + 2\n(\n-2) \ga_{\Y} = 0 \,, \nonumber\\
   &&\hspace*{-2cm} n\ga_W - \n\ga_{\W}
   + 2(n\ga_Y - \n\ga_{\Y}) = 0 \,, \qquad  \ga_V = 0 \,,
\eee
for $n\neq 1$, $\n\neq 1$. For $n=1$ or $\n=1$ $\Phi_2^{n,\n}$
is closed as a consequence of (\ref{nonzero}).

Our problem is to investigate whether there exist coefficients
$R_1$, $W_1$, ... such that $R_2^{\cal D}$, $W_2^{\cal D}$, ...
have a form (\ref{a:t}). To this end one has to solve the system
(\ref{^R})-(\ref{^Y}) in terms of the formal series
$f(\tau) = \sum_{k=-\infty}^{p<\infty} f_k \tau^k$.

The Fierz transformations (\ref{delta}) for $\Phi_1$ together
the exact shifts of $R_1$, $W_1$, ... by any $R_1^{\cal D}$,
$W_1^{\cal D}$, ... (\ref{^phi1}) produce the following equivalence
transformations
\bee
\label{de}
   && \delta R_1 = -\dt \chi_1 + \gvep \,,\qquad
   \delta W_1 = -\dt \xi_1 + 2i\n \chi_1 - 2(1-i\tau)\gvep \,,
   \nonumber \\
   && \delta \W_1 = -\dt \bxi_1 - 2in \chi_1 - 2(1+i\tau)\gvep\,,
   \qquad
   \delta V_1 = 2(1+\tau^2)\gvep - i(n\xi_1 - \n\bxi_1)\,,
   \nonumber\\
   && \delta Y_1 = i(\n-1)\xi_1  \,,\qquad
   \delta \Y_1 = -i(n-1)\bxi_1  \,.
\eee

We consider separately two cases: (i) with  $n=1$, $\n =1-2s$
or $\n=1$, $n =1-2s$ and (ii) with  $n < 1$ and $\n < 1$.
As shown below, the case (i) corresponds to the nontrivial physical
conserved currents, whereas the case (ii) describes all possible
``improvements''.

Let us start with the case (i) setting for definiteness $\n=1$.
According to (\ref{nonzero}), $Y_2$ is the only coefficient giving
a non-zero contribution to $\Phi^{n,1}_2(y)$. Obviously, a 2-form
with $\n=1$ is invariant under the transformations (\ref{delta}).
The only non-trivial equation is (\ref{^Y}). With $Y_2^{\cal D}$
(\ref{a:t}) it takes the form
\be
\label{dY}
   (1+\tau^2)\; \dt Y_1 = -i(1+i\tau)\;Y_1
   + i(2s-1)(1-i\tau)\;Y_1 + i\ga_Y(1-2s,1)\; z^{1-2s}\z\, \tau^{-1}
\ee
modulo terms polynomial in $\tau$. It is not hard to see that
a generic solution of (\ref{dY}) is
\bee
\label{Y(tau)}
    Y_1(z,\z,\tau)  = -{i\over 2}\; \ga_Y(1-2s,1) z^{1-2s} \z \;
    (1-i\tau) (1+i\tau)^{2s-1} \ln{(1+\tau^{-2})}
       \nonumber \\
    {}+ \gs \; z^{1-2s}\z \; (1-i\tau) (1+i\tau)^{2s-1} \;
    \ln{\left({1+i\tau^{-1}\over 1-i\tau^{-1}} \right)}+Q(\tau)\,,
\eee
where $\gs$ is an arbitrary constant and $Q(\tau)$ is some
inessential polynomial. The logarithms are treated as power series
in $\tau^{-1}$.

At any $\gs$, the solution (\ref{Y(tau)}) is an infinite series
in $\tau^{-1}$, thus corresponding to some pseudolocal 1-form.
Thus, the 2-forms $\Phi^{s}_2(y|x)$ constructed with the
polynomials $Y_2$ at $\n=1$ and with $\Y_2$ at $n=1$ are
${\cal D}$-closed and cannot be represented as $D\Phi^{s}_1(y|x)$
with a spacetime local $\Phi^{s}_1(y|x)$. We therefore argue that
the 2-form $\Phi^{s}_2(y)$ describes a physical conserved
current of spin $s$. The currents (\ref{s-scal}) as well as
the currents containing fermions are reproduced via $Y_2$ (\ref{a:t})
with $\ga_Y(1-2s,1) = 2^{2s-1}\;(\gl\psi)^{[s]}$.
The formula (\ref{Y(tau)}) solves the problem of reformulation of
the physical currents as pseudolocally exact 2-forms.
Note that the currents generated by $Y_2(n=1-2s,\n=1)$ and
$\Y_2(n=1,\n=1-2s)$ are equivalent by the interchange
$C\leftrightarrow C'$.

Note that the solution (\ref{Y(tau)}) is not unique, containing
an arbitrary parameter $\gs$. Since
the transformations (\ref{de}) are trivial for $Y_1$
at $\n=1$, this one-parametric ambiguity cannot be compensated
this way. This means that we have found a pseudolocal 1-form that
is ${\cal D}$-closed but not ${\cal D}$-exact, i.e., the cohomology
group $H^1 (T,{\cal D})$ is nontrivial.
This fact is in agreement with the one-parametric ambiguity found
in \cite{PV} for the spin 2 case.

Let us now consider the case of $n<1$, $\n<1$.
Substituting (\ref{a:t}) into the system (\ref{^R})-(\ref{^Y})
one can see that, if the conditions (\ref{closed}) guaranteeing that
$\Phi_2$ is ${\cal D}$-closed are satisfied, then it admits
the following solution,
\bee
\label{Rt}
    &&\hspace*{-1.5cm} R_1(z,\z,\tau) = {1\over 4n\n}
    \left( n\ga_W + \n\ga_{\W} - {2n\over \n-1}\, \ga_Y
    - {2\n\over n-1}\, \ga_{\Y} \right)\, z^n \z^{\n} \tau^{-1} \,,
    \quad  n,\n<0\,, \\
\label{Wt}
    &&\hspace*{-1.5cm} W_1(z,\z,\tau) =
    {\ga_{Y}\over \n-1}\, z^n \z^{\n} \tau^{-1}\,,\quad
    \W_1(z,\z,\tau) = {\ga_{\Y}\over n-1}\,
    z^n \z^{\n} \tau^{-1} \,,\quad
     V_1 = Y_1 = \Y_1 = 0 \,.
\eee
We observe that the 1-form $\Phi^{n,\n}_1(y)$
leads to a spacetime local expression since $R_1$, $W_1$,
and $\W_1$ (\ref{Rt}), (\ref{Wt}) are linear in $\tau^{-1}$.
Therefore, $\hat{\Phi}^{n,\n}_2(y|x) = D\Phi^{n,\n}_1(y|x)$
with some local $\Phi^{n,\n}_1(y|x)$. Thus, it is an ``improvement''
of the physical current 2-form $J(C^2)$ on the r.h.s. of
(\ref{Ch-S}), which can be compensated by a local field
redefinition of the (HS) gauge fields.

To investigate what happens in the flat limit $\gl\to 0$
one should use the generating function $\tilde{C}(y)$ (\ref{Cy})
instead of $C(y)$. A simple analysis show that in terms of
$\tilde{C}(y)$ the solution (\ref{Y(tau)}) contain an inverse power
of $\gl$ together with each power of $\tau^{-1}$.
Hence, a representation of physical current 2-forms $\Phi^{s}_2(y)$
as some differentials $D\Phi^{s}_1(y)$ becomes meaningless in the flat
limit.

\section*{Conclusion}

It is shown that local conserved currents of an arbitrary spin
in $AdS_3$ can be treated as ``improvements" within the class of
infinite power expansions in higher derivatives, i.e., 2-forms $J$
dual to the physical conserved currents are shown to be exact in
this class, $J=DU$. The 1-forms $U$ are
constructed explicitly what allows us to write down nonlocal field
redefinitions compensating matter sources in the equations of motion
for the Chern-Simons gauge fields of all spins.
The coefficients in the expansion of $U$ in derivatives of the
matter fields contain negative powers of the cosmological constant
(i.e. positive powers of the AdS radius) and therefore do not admit
a flat limit. The existence of $U$ may be related to the holography
in the AdS/CFT correspondence since it indicates that
local current interactions in $AdS_3$ are in a certain sense
trivial and can, up to some surface terms, be compensated by a
field redefinition. We expect similar phenomenon to take place for
$AdS_d$ with any $d$.

\bigskip
\noindent
{\bf Acknowledgments.} This research was supported in part by
INTAS Grant No.96-0538 and by the RFBR Grants No.99-02-16207
and No.96-15-96463.
S.~P. acknowledges a partial support from the Landau Scholarship
Foundation, Forschungszentrum J\"ulich.

\end{document}